\documentclass[prb,raggedbottom,twocolumn,superscriptaddress,floatfix,amsmath,A4]{revtex4}
\usepackage[latin1]{inputenc}
\usepackage[english]{babel}
\usepackage[T1]{fontenc}

\usepackage{graphicx}
\usepackage{amssymb}
\usepackage{amsmath}
\usepackage{amscd}
\usepackage{eucal}
\usepackage{color}
\usepackage{bm}

\newcommand{\e}{{\rm e}}
\newcommand{\rmd}{{\rm d}}

\newcommand{\half}{{\textstyle{\frac{1}{2}}}}

\newcommand{\eps}{\epsilon}

\newcommand{\kB}{k_{\rm B}}

\definecolor{DarkGreen}{rgb}{0,0.7,0}

\begin{document}

\title{
Nonlinear thermoelectricity in point-contacts at pinch-off: 
a catastrophe aids cooling}

\author{Robert S.~Whitney}
\affiliation{
Laboratoire de Physique et Mod\'elisation des Milieux Condens\'es (UMR 5493), 
Universit\'e Grenoble 1, Maison des Magist\`eres, B.P.~166, 38042 Grenoble, France.}

\date{August 30, 2012 --- Last revision July 18, 2013}
\begin{abstract}
I consider refrigeration 
and heat engine circuits based on the nonlinear thermoelectric response of 
point-contacts at pinch-off, allowing for electrostatic interaction effects.
I show that a refrigerator can cool to much lower temperatures than predicted 
by the thermoelectric figure of merit $ZT$ 
(which is based on linear-response arguments).
The lowest achievable temperature has a discontinuity,
called a {\it fold catastrophe} in mathematics, at a critical driving current $I=I_{\rm c}$. 
For $I >I_{\rm c}$ one can in principle cool to absolute zero,
when for $I<I_{\rm c}$ the lowest temperature is about half the ambient temperature. 
Heat back-flow due to phonons and photons stop cooling at a temperature above absolute zero, 
and above a certain threshold turns the discontinuity into a sharp cusp.
I also give a heuristic condition for when an arbitrary 
system's nonlinear response means that its $ZT$ ceases to indicate (even qualitatively)
the lowest temperature to which the system can refrigerate.
\end{abstract}
\maketitle

\section{Introduction}
Nanostructures often have thermoelectric responses,  
with electrical-currents causing heat-currents, and vice-versa 
\cite{books,DiSalvo-review,Shakouri-reviews}. 
There have recently been a number of proposals for nanostructures or molecules 
with large thermoelectric responses  
\cite{Casati2008,Nozaki2010,Saha2011,Wierzbicki2011,Karlstrom2011,Gunst2011,Rajput2011,Trocha2012}
which could have engineering applications for efficient thermoelectric power-generation and refrigeration.
In particular, it is hoped that they could cool electrons well below the 
temperature of standard cryostats 
\cite{Pekola-reviews,SC-cooling-expt1,SC-cooling-expt2,SC-cooling-expt3}, 
which are increasingly inefficient at sub-Kelvin temperatures.

However, {\it good} nanostructure refrigerators (those which cool to significantly below their environment's temperature) are rarely in the linear-response regime.
Linear-response theory works for small temperature drops (compared with the average temperature) at the scale of the nanostructure
and the scale of the electron's inelastic scattering length.
This is often the case in bulk semoconductors \cite{Footnote:semicond,Zebarjadi2007}, but {\it not} in such nanostructures.  See, for example, experiments on refrigeration with
S-N tunnel junctions,
that generate a temperature drop from 300mK to 100mK
across a tunnel junction \cite{SC-cooling-expt1,SC-cooling-expt2,SC-cooling-expt3,Pekola-reviews}.

\begin{figure}[b]
\includegraphics[width=\columnwidth]{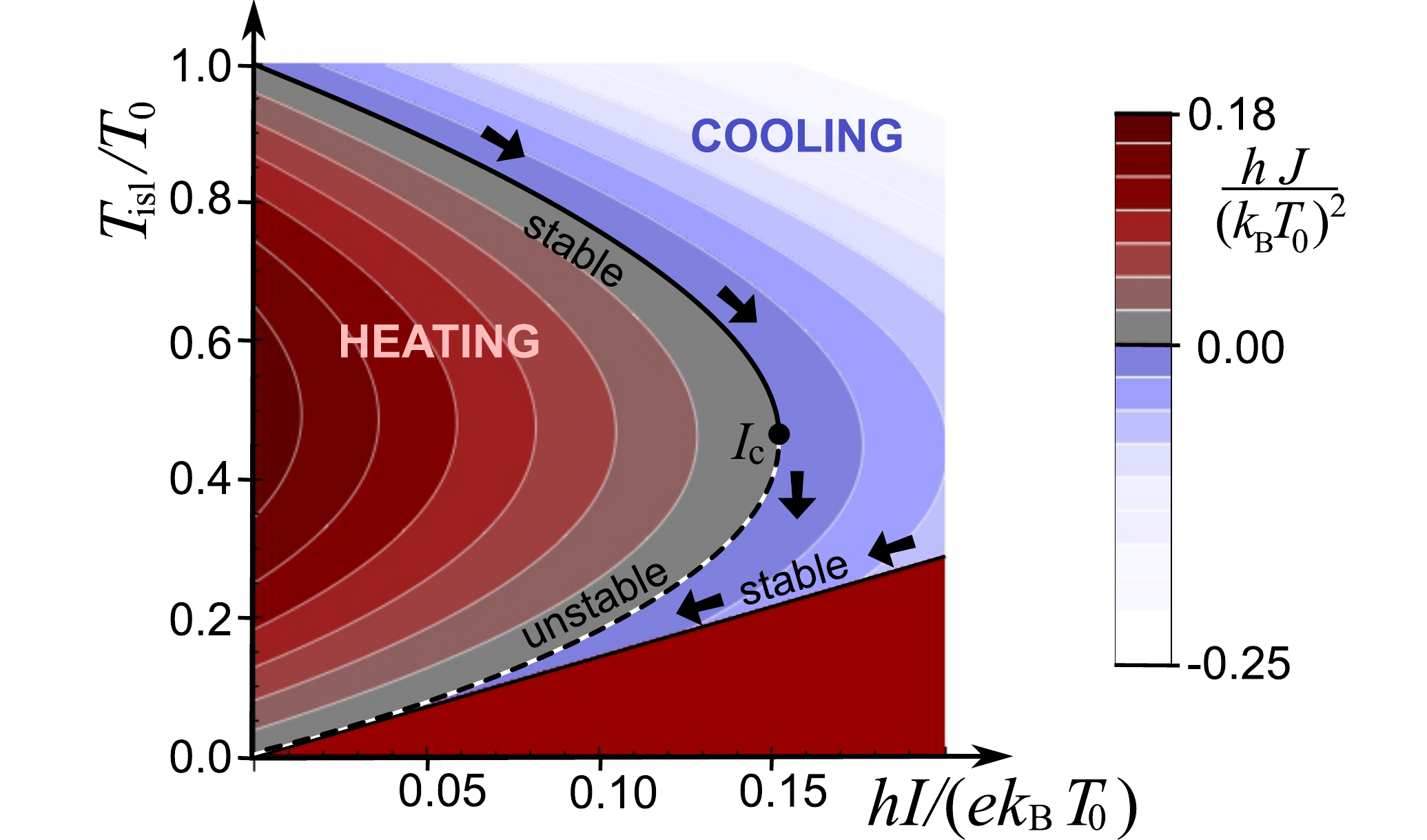}
\caption{\label{Fig:Cooling-no-phonons}
Heat-current $J(T_{\rm isl},I)$ through a point-contact when driven with a current $I$,
for negligible phonon or photon heating.
Blue indicates cooling of the island in Fig.~\ref{Fig:circuit}a, while red indicates heating.
The solid curve is the steady-state ($J=0$), with the catastrophe at $I_{\rm c}$.
The straight line is the maximum current, $I_{\rm max}$, corresponding to infinite bias.
}
\end{figure}

Unfortunately, there is no general theory for the {\it nonlinear} response of quantum systems,
because interaction effects are usually significant,
and must be modeled using approximations appropriate for the system in question.
Here,
I calculate the {\it fully} nonlinear thermoelectric response of a point-contact
at pinch off.  This system is one of the main candidates for a nanoscale thermoelectric, and its linear (and nearly linear) thermoelectric response is well-studied 
experimentally \cite{Molenkamp1992,Ghoshal2002} and theoretically
\cite{Molenkamp1992,Bogachek1998,Cipiloglu2004,Nakpathomkun-Hu-Linke2010}.
I consider this thermoelectric response when the temperature drop across the 
point-contact is of order the average temperature,
for which the response is far outside its linear regime. 
This can be modeled with a nonlinear Landauer-B\"uttiker scattering 
theory~\cite{Moskalets1995,Christen-ButtikerEPL96,Sanchez-Buttiker,
Meair-Jacquod2012,Sanchez-Lopez2012} 
for thermoelectric heat-transport~\cite{Butcher1990}.
I find that
the dimensionless figure of merit, $ZT$, ceases to be a good measure of the 
thermoelectric response outside the linear regime. 
Electricity generation is {\it worse}
than linear-response theory indicates,
but refrigeration is {\it better} (achieving much lower
temperatures than linear-response theory predicts).
Indeed, the lowest temperature of the refrigerator is a discontinuous function of the electrical current. This discontinuity --- a fold catastrophe in mathematical language ---
occurs at a critical current $I_{\rm c}$, and helps refrigeration.
For currents $I < I_{\rm c}$
the refrigerator cannot cool below a finite temperature (about half the ambient temperature
for $I \to I_{\rm c}$),
while for $I>I_{\rm c}$ it passes the catastrophe
and can {\it in principle} cool to absolute zero (see Fig.~\ref{Fig:Cooling-no-phonons}).

In practice a thermoelectric device's quality is reduced by the nonlinear
back-flow of heat carried by chargeless particles; phonons and photons.
When such back-flow effects are weak, 
the catastrophe is little affected, 
but cooling stops at a temperature above absolute zero. 
At a critical value of back-flow effects, 
the catastrophe becomes a cusp (discontinuity in the derivative) of the dependence of the lowest temperature on $I$.  
The nonlinear nature of the cusp still means the lowest achievable temperature is lower than linear theory predicts.

\begin{figure}
\includegraphics[width=\columnwidth]{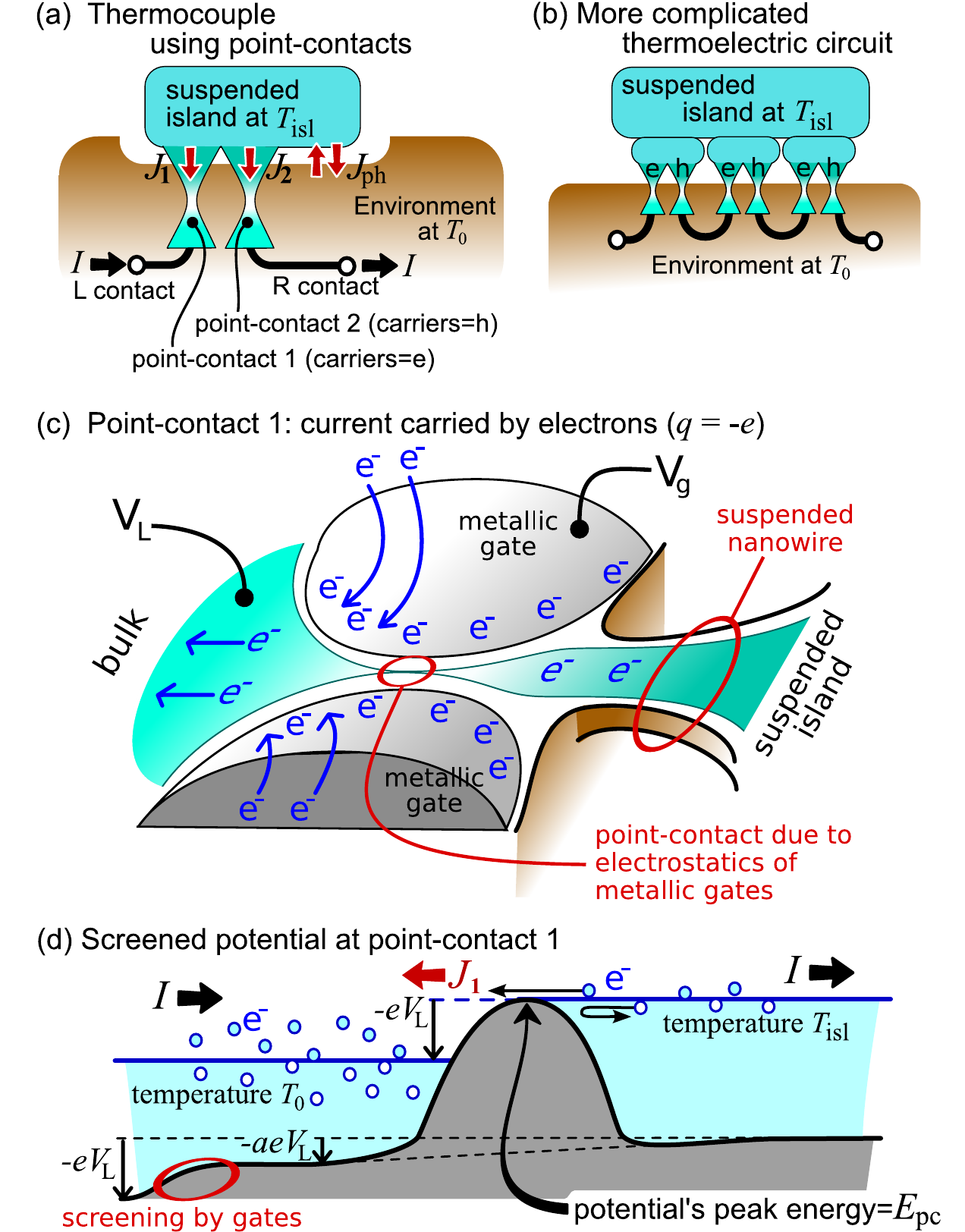}
\caption{\label{Fig:circuit}
Thermoelectric circuits made with point-contacts shown in (a,b);
``e'' (``h'') means the point-contact is in a material whose charge-carriers are electrons (holes).
One should minimize the heat-current carried by phonons and photons,  $J_{\rm ph}$, 
by suspending the island \cite{suspended-structures,suspended-structures2}.
The temperature of the island in similar set-ups 
(albeit not suspended) has been probed experimentally using a quantum dot as thermometer\cite{Gasparinetti2011}, 
although not yet in the regime of refrigeration.
(c) Motion of charges in the gates (arrows) caused by making $V_{\rm L}$ positive, 
which partially screens $V_{\rm L}$ at some distance from the point-contact.
(d) Point-contact 1 tuned to pinch-off ($E_{\rm pc}$ equals the 
island's chemical potential) by adjusting $V_{\rm g}$.
}
\end{figure}

\section{Nearly-linear analysis for any system,
and its breakdown} 
\label{Sect:breakdown}

The usual ``nearly-linear'' analysis \cite{books} 
takes linear response theory plus a Joule heating term,
and enables one to quantify devices in terms of their dimensionless figure of merit
$ZT= G\Pi^2\big/\big((\Theta_{\rm el}+\Theta_{\rm ph})T\big)$, 
where $T$ is the device temperature, 
$\Pi$ is its Peltier coefficient, $G$ and $\Theta_{\rm el}$ are 
electrical and thermal conductances of electrons, while 
$\Theta_{\rm ph}$ is the thermal conductivity of chargeless excitations, principally phonons and photons.  
This nearly-linear analysis predicts electric power generation (when the island is heated) with an efficiency
\begin{eqnarray}
\eta &=& {\sqrt{ZT+1} -1 \over \sqrt{ZT+1} +1} \left( 1 - {T_0\over T_{\rm isl}} \right) ,
\label{Eq:Eff-linear}
\end{eqnarray}
where $T_{\rm isl}$ and $T_0$ are the island and environment temperatures.
Typically, $ZT$ is taken at the temperature $\sim\half(T_0+T_{\rm isl})$.
Carnot efficiency corresponds to  $ZT \to \infty$.
For refrigeration, it predicts that the lowest achievable temperature, $T_{\rm min}$, is given by
\begin{eqnarray}
T_{\rm min}/ T_0 &=&  1 - \half ZT.
\label{Eq:Tmin-linear}
\end{eqnarray}
Eq.~(\ref{Eq:Tmin-linear}) is derived \cite{books} by  combining linear response terms
(the Peltier effect due to the current $I$, and heat flow due to the temperature difference,
$T_0-T_{\rm isl}$), with a nonlinear $I^2$ term corresponding to Joule heating. 
Heat flow out of the island in Fig.~\ref{Fig:circuit}a, due to a current $I$ passing though
element 1, then the island and then element 2,
is 
\begin{eqnarray}
J(T_{\rm isl}, I) \ \simeq\   \Pi_- I \ -\  \Theta_+ \,(T_0-T_{\rm isl})  \, -\, \half G_+^{-1}  I^2,
\label{Eq:J-nearly-linear}
\end{eqnarray}
for $\Pi_-=\Pi_2 -\Pi_1$, $\Theta_+=\Theta_1+\Theta_2+\Theta_{\rm ph}$
and $G_+^{-1} = G_1^{-1}+G_2^{-1}$.
Here $\Pi_i$, $G_i$ and $\Theta_i$ are the Peltier coefficient, 
the electrical and thermal conductances of element $i$. 
The steady-state curve, $J=0$, gives $T_{\rm isl}$ as a quadratic function of $I$. 
The parabola's minimum is $T_{\rm min}$ in  Eq.~(\ref{Eq:Tmin-linear}) with $ZT= G_+\Pi_-^2/(\Theta_+ T)$.

For a point-contact at pinch-off (see Section~\ref{Sect:nonlinear}),
linear-response
\cite{Engquist-Anderson1981,Sivan-Imry1986,Butcher1990,Claughton-Lambert,
jw-epl,Liu2011,jwmb-onsager} gives
$G_1 = (e^2/h) \, (1 /2)$, 
$\Pi_1 = -(k_{\rm B}T_0/ e) \, 2\ln(2)$,
and\cite{footnote:Theta} $ \Theta_1 = (k_{\rm B}^2 T_0/h)\, \big( \pi^2/ 6 \,-\, 2[\ln(2)]^2 \big)$. 
Thus $ZT \simeq 1.4$ so
\begin{eqnarray}
\eta = 0.22\,(1-T_{\rm 0}/T_{\rm isl}),     \qquad \qquad T_{\rm min}= 0.3 \,T_0.
\label{Eq:linear-results}
\end{eqnarray}

However, Eq.~(\ref{Eq:J-nearly-linear}) ceases to apply whenever thermoelectric effects 
are strong enough that the 
nonlinear terms that were not included in Eq.~(\ref{Eq:J-nearly-linear}) become relevant.
Heuristically, 
Eq.~(\ref{Eq:J-nearly-linear}) fails to get the physics {\it qualitively} correct for any system
where the nonlinear Peltier term \cite{Cipiloglu2004}, $\widetilde{\Pi}\, I^2$, 
is larger than the Joule heating term $\half G_+^{-1} I^2$.
The reason being that including $\widetilde{\Pi}\, I^2$ in Eq.~(\ref{Eq:J-nearly-linear}) then changes the 
sign of the prefactor on $I^2$;
so one must go beyond $I^2$ to find
the steady-state curve's minimum. 
Including this $\widetilde{\Pi}$ term will then make the refrigerator  {\it better} than if it were neglected.
However higher order terms ($I^3$ or higher) will then also be crucial in determining the 
lowest temperature the refrigerator can achieve. 

This break-down of the nearly-linear theory as $\widetilde{\Pi}$ increases is 
discussed in Section~\ref{sect:composite}  for a particular system 
(a point-contact in parallel with a tunnel-barrier) in which  
$\widetilde{\Pi}$ can be varied; it indeed occurs when $\widetilde{\Pi}$ is of order $\half G_+^{-1}$.
Readers familiar with simple $\phi^4$-theory will see this is
similar to the para- to ferro-magnetic crossover 
of a magnet in a B-field upon reducing the temperature \cite{footnote:ferro-para}.
However, unlike in $\phi^4$-theory, for small barrier transmission (or a point-contact alone),
the analysis shows such strong nonlinearities (catastrophe, etc) that the minimum is not captured by a perturbative expansion in $I$   
up to any order.

Of course, the above heuristic argument assumes that
nonlinearities in the thermoelectric response are significant when the linear thermoelectric response is significant.  While in most cases this is true, the S-N tunnel junction is a 
counter-example;
it has no thermoelectric response in the linear regime (so $ZT=0$), 
but does have a large nonlinear response which has been used for 
refrigeration \cite{SC-cooling-expt1,SC-cooling-expt2,SC-cooling-expt3}. 
This is because the electron and holes have the same transmission at zero bias
(so there is no thermoelectric response), but nonlinear charging effects
enhance the transmission of electrons over those of holes creating an entirely nonlinear
thermoelectric effect. 
It would be interesting to see if such S-N junctions exhibit the type of
catastrophes found in this work for point contacts.

\section{Fully nonlinear analysis for point-contacts}
\label{Sect:nonlinear}
The thermoelectricity literature \cite{books,DiSalvo-review,Shakouri-reviews} 
discusses $J(T_{\rm isl},I)$ 
--- as in Eq.~(\ref{Eq:J-nearly-linear}) above --- rather than $J(T_{\rm isl},V)$ for voltage drop, $V$.
This is because different thermoelectric devices are arranged in series electrically 
(see Fig.~\ref{Fig:circuit}a,b), so $I$ is the same in all of them (unlike voltage drops).
Thus it is easier to get response of a
series of elements from each element's $J(T_{\rm isl},I)$ than from each element's $J(T_{\rm isl},V)$. 
For complicated non-linear responses, the former is straight-forward while the latter is extremely difficult; thus I  consider $J(T_{\rm isl},I)$.

I  take the island to be classical; i.e.~ big enough for particles entering it
to thermalize to a Fermi distribution at temperature $T_{\rm isl}$ before escaping.
I also assume quantum charging effects (Coulomb blockade, etc) 
within the island are negligible,
while classical charging effects enure electro-neutrality
(i.e. that the sum of electrical currents into the island is zero).
Refs.~[\onlinecite{Entin-Wohlman2010,SSJB2012,Sothmann-Buttiker2012,Sanchez-Buttiker2012,Entin-Wohlman2013}] 
consider cases where
there is a quantum dot in place of the classical island. 
In our case, each point-contact can be treated by a 
separate Landauer-B\"uttiker scattering matrix analysis
\cite{Butcher1990}, see also Refs.~[\onlinecite{Engquist-Anderson1981,Sivan-Imry1986,
Claughton-Lambert,jw-epl,Liu2011,jwmb-onsager}].
I  generalize these heat currents to the nonlinear regime \cite{w-joule},
including electrostatic (Hartree-like) interaction effects in a self-consistent and gauge-invariant manner;
as Refs.~[\onlinecite{Christen-ButtikerEPL96,Sanchez-Buttiker,Meair-Jacquod2012,Sanchez-Lopez2012}]
did for charge-current, 
see also \cite{Buttiker1993-96,Petitjean2009}.
To go beyond the voltage-squared contributions to transport
(which Ref.~[\onlinecite{Christen-ButtikerEPL96}] treated in detail), 
I  use a simple model of interactions, 
which is none the less gauge-invariant and self-consistent.
The charge-current, $I_i$, and heat-current, $J_i$,  into lead $i$ of a given nanostructure 
are 
\begin{eqnarray}
I_i \!&=&\!  -\int_{-\infty}^\infty {{\rm d}\eps \over h} \sum_j \,q\,   
{\cal A}_{ij}\big( \{\eps -qV_k\} \big) \  f_j(\eps)  ,
\label{Eq:I-initial}
\\
J_i \!&=&\!  -\int_{-\infty}^\infty {{\rm d}\eps \over h} \sum_j \, (\eps- qV_i) \  
{\cal A}_{ij}\big( \{\eps -qV_k\} \big) \  f_j(\eps)  ,\ \ \ 
\label{Eq:J-initial}
\end{eqnarray}
where $f_j(\eps) = \big( 1+ \exp\big[(\eps-qV_j) \big/ (\kB T_j) \big] \big)^{-1}$ is the Fermi function, 
and $q$ is the charge of the carriers; 
electrons with $q=-e$ in point-contact 1 and holes with $q=e$ in point-contact 2.
The energy $\eps$ and all voltages $V_k$ are measured from the same external reference.
The transmission function of a particle 
through the nanostructure from lead $j$ to lead $i$
is 
${\cal A}_{ij}\big( \{\eps -qV_k \} \big)
= {\rm Tr} \left[ {\bf 1}_i\delta_{ij} - 
{\cal S}^\dagger_{ij}\big( \{\eps -qV_k \} \big) 
\,{\cal S}_{ij}\big( \{\eps -qV_k \} \big)\right] $,
where  ${\cal S}_{ij}$ is the scattering matrix from lead $j$ to lead $i$, and the trace is over all modes of those leads.
Here ${\cal S}_{ij}$ must be found {\it self-consistently}; 
it depends on the charge distribution in the nanostructure, which in turn depends on ${\cal S}_{ij}$. 
Writing ${\cal S}_{ij}$ as a function only of energy differences,  $\{\eps -qV_k \}$,
makes the gauge-invariance explicit;
it satisfies \cite{Christen-ButtikerEPL96}
$\big[\big(\rmd /\rmd \eps \big) + \sum_k \big( \rmd /\rmd  (qV_k)\big)\big]{\cal A}_{ij} =0$.

Point-contact 1 is a two-lead nanostructure with electron charge-carriers
($q=-e$).  The gauge-invariance means one is free to measure all
energies $\eps$ and voltages $V_k$  (including $V_{\rm g}$) from
the island's chemical potential (the point-contact's M lead).
I assume that a proportion $(1-a)$ of $V_{\rm L}$ is screened by the electrostatic gates 
a long way from the narrowest-point of the point-contact, while
the rest is screened self-consistently by the electron-gas (Fig.~\ref{Fig:circuit}d)
close to the point-contact.
Then the {\it screened} point-contact induces a potential barrier of height, $E_{\rm pc}$ (measured from the island's chemical potential), typically obeying
$E_{\rm pc} -  E_{\rm g}= E_{\rm scr}(a qV_{\rm L})$,
where $E_{\rm scr}$ is due to screening. 
Here $E_{\rm g}$ can be tuned at will, since it is $eV_{\rm g}$ minus a 
geometry-dependent constant.
Assuming a long enough point-contact that there is negligible
tunnelling, one has
${\cal A}_{\rm LM}(\eps-E_{\rm pc}>0)=-1$ (perfect transmission) and  
${\cal A}_{\rm LM}(\eps-E_{\rm pc}<0)=0$ (no transmission) \cite{footnote:tunnelling}.
As an example, the appendix gives
a simple screening model for which I 
derive $E_{\rm scr}(a qV_{\rm L})$ self-consistently. 
However, in what follows I 
allow the nature of screening (both $a$ and the form of $E_{\rm scr}(a qV_{\rm L})$) 
to be completely arbitrary. 

For any given $V_{\rm L}$, one can adjust $V_{\rm g}$ to tune to pinch-off  ($E_{\rm pc}=0$).
If the gates dominate screening ($a \to 0$), then $E_{\rm pc}$ is
$V_{\rm L}$-independent, making this straightforward.
Otherwise, the point-contact should be calibrated prior to use; finding the pinch-off point
(the $V_{\rm g}$ at which current starts to flow), as a function of $V_{\rm L}$.  
At pinch-off, the currents from point-contact 1 into the island are 
\begin{eqnarray}
I (T_{\rm isl},V_{\rm L}) \! \! &=& \!{e \kB\over h} \left[T_{\rm isl} \ln(2) -  T_0 \ln \left( 1+ \e^{-eV_{\rm L}/\kB T_0} \!\right)\right] \! , \ \ \ 
\label{Eq:IvsV}
\\
J_1(T_{\rm isl},V_{\rm L}) \!\! &=& \! -{\kB^2 \over h}  \left[ T_{\rm isl}^2 {\pi^2\over 12}  + T_0^2 
{\rm Li}_2 \left(-\e^{-eV_{\rm L}/\kB T_0} \right)\right] \! , \qquad
\label{Eq:J_1vsV} 
\end{eqnarray}
where ${\rm Li}_2(z)$ is a dilogarithm function. 
Eqs.~(\ref{Eq:IvsV},\ref{Eq:J_1vsV}) give
\begin{eqnarray}
J_1(T_{\rm isl},I) = -{\kB^2T_0^2 \over h}  \Bigg[ {\pi^2 T_{\rm isl}^2\over 12T_0^2}  
+  {\rm Li}_2\Big(1-\exp \left[ {\cal I}(T_{\rm isl},I) \right]  \! \Big) \Bigg] \! ,
\label{Eq:J1-vs-I}
\end{eqnarray}
where I define 
${\cal I} = h\big[I_{\rm max}(T_{\rm isl})-I\big]\big/(e\kB T_0)$ and note that 
$I \leq I^{\rm max} (T_{\rm isl} )= e\kB T_{\rm isl} \ln[2]/h$.
This function is given by the color plot 
in Fig.~\ref{Fig:Cooling-no-phonons}.
For point-contact 2 
(where carriers are holes not electrons) one takes $-e \leftrightarrow e$, then  
$J_2(T_{\rm isl},I)=J_1(T_{\rm isl},I)$ since  $I_2=-I$.

For 
${\cal I}  \ll 1$, one can use ${\rm Li}_2(z) = z +{\cal O}[z^2]$ to
write
\begin{eqnarray}
J_1 = \big(\kB^2T_0^2/h \big)  \big[ {\cal I}-(\pi^2/12) (T_{\rm isl}/T_0)^2
+{\cal O}[{\cal I}^2]
 \big],
\label{Eq:J_1approx}
\end{eqnarray}
so $J_1$ as a quadratic in temperature and linear in current; the reverse of
the nearly-linear theory in Eq.~(\ref{Eq:J-nearly-linear}).
This approximation
captures the features of the exact result plotted in Fig.~\ref{Fig:Cooling-no-phonons}, 
except the top-left corner.
This corner is the linear-response regime 
(small $(T_0-T_{\rm isl})$ and $I$), where one has Eq.~(\ref{Eq:J1-vs-I})
with
${\rm Li}_2(-1+z) \simeq -\pi^2/12\,+\,\ln[2] z$.

\section{Refrigeration without phonons or photons}
Heat flow into the island is
$J_{\rm total} \propto J_1$ for the devices in Fig.~\ref{Fig:circuit}a,b; $J_{\rm total}=2J_1$ for the thermocouple.
The black curves in Fig.~\ref{Fig:Cooling-no-phonons}
are $J_{\rm total}=0$, 
giving the steady-state temperature
(solid for stable steady-states and dashed for unstable ones). Solid curves
give the temperature the island will be cooled to by a current $I$.
Eq.~(\ref{Eq:J_1approx}) tells us the steady-state 
has $I$ as a quadratic function of 
$T_{\rm isl}$;  this approximation gives the catastrophe at 
$eI_c/(e\kB T_0)=3(\ln[2]/\pi)^2\simeq 0.14$ with $T_{\rm isl}/T_0 = 6 \ln[2]/\pi^2 \simeq 0.42$,
which is very close to the exact solution in Fig.~\ref{Fig:Cooling-no-phonons}.

\begin{figure}
\includegraphics[width=\columnwidth]{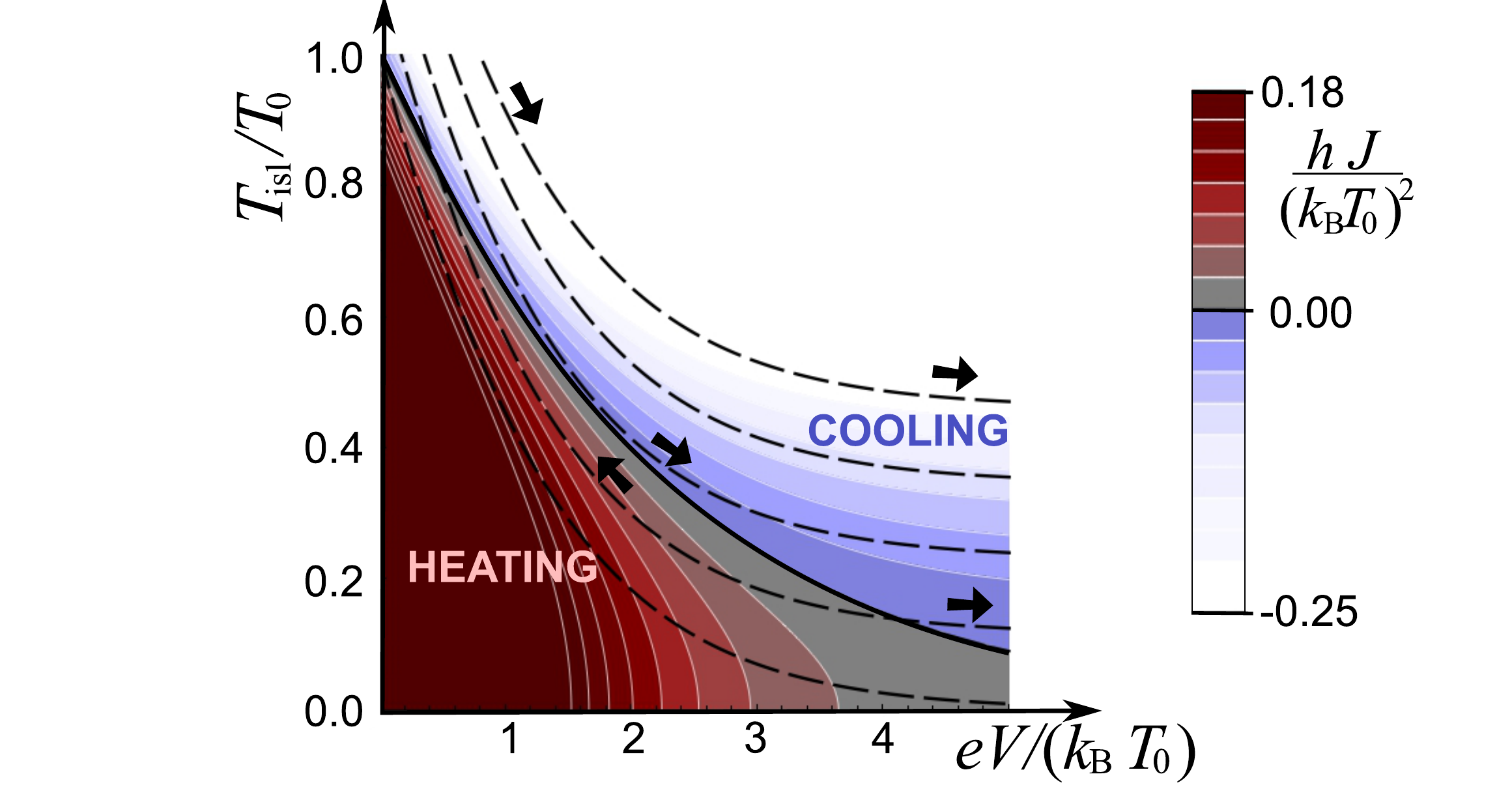}
\caption{\label{Fig:Cooling-V}
Heat-current as a function of $V$ and $T_{\rm isl}$, with the steady-state, $J=0$, marked by the black-curve
(note the darkest color is used for all $hJ/(\kB T_0)^2 >0.18$ and white for all 
$hJ/(\kB T_0)^2 <-0.25$).
Superimpose are lines of constant current (dashed),  these are $hI/(e\kB T_0) =$ 0, 0.08, 0.16, 0.24, 0.32 from bottom to top.  
}
\end{figure}

\subsection{Voltage dependence of cooling.} 
As mentioned above, one typically consider
the response of thermoelectric devices as a function of current $I$, rather than
voltage $V$, since $I$ is conserved in series electrical circuits like those in Fig.~\ref{Fig:circuit}a,b.
However the origin of the catastrophe can be seen in Fig.~\ref{Fig:Cooling-V}, which compares the 
nonlinear response of the point-contact as a function of voltage, Eq.~(\ref{Eq:J_1vsV}), with curves of 
constant current given by Eq.~(\ref{Eq:IvsV}).
Curves with  $I<I_{\rm c}$, such as $hI/(e\kB T_0) =0.08$, cross from cooling to heating and back again, while curves with  $I>I_{\rm c}$, such as $hI/(e\kB T_0) =0.16,0.24, 0.32$,
never enter the heating regime.  For larger $I$, the temperature saturates at a higher value.  All of this fits with Fig.~\ref{Fig:Cooling-no-phonons}.

Note that if one wants the voltage response of  the circuits in Fig.~\ref{Fig:circuit}a,b
one cannot easily get if from the voltage response of each element, as plotted in Fig.~\ref{Fig:Cooling-V}. 
This is because the voltage drop across each thermoelectric element 
depends nonlinearly on $T_{\rm isl}$, even when the total voltage drop across all elements is fixed
(unless all thermoelectric elements have 
the same $I(T_{\rm isl},V)$).  Indeed the simplest way to get the voltage response of  the circuits in 
Fig.~\ref{Fig:circuit}a,b is to take the heat-flow in each circuit element {\it as a function of $I$} 
(as plotted in Fig.~\ref{Fig:Cooling-no-phonons}). 
Since $I$ is the same in every element, one can get the heat-flow as a function of  the voltage drop 
across the circuit as the sum of voltage drops across each element
(as a function of $T_{\rm isl}$ and $I$).

\begin{figure}
\includegraphics[width=\columnwidth]{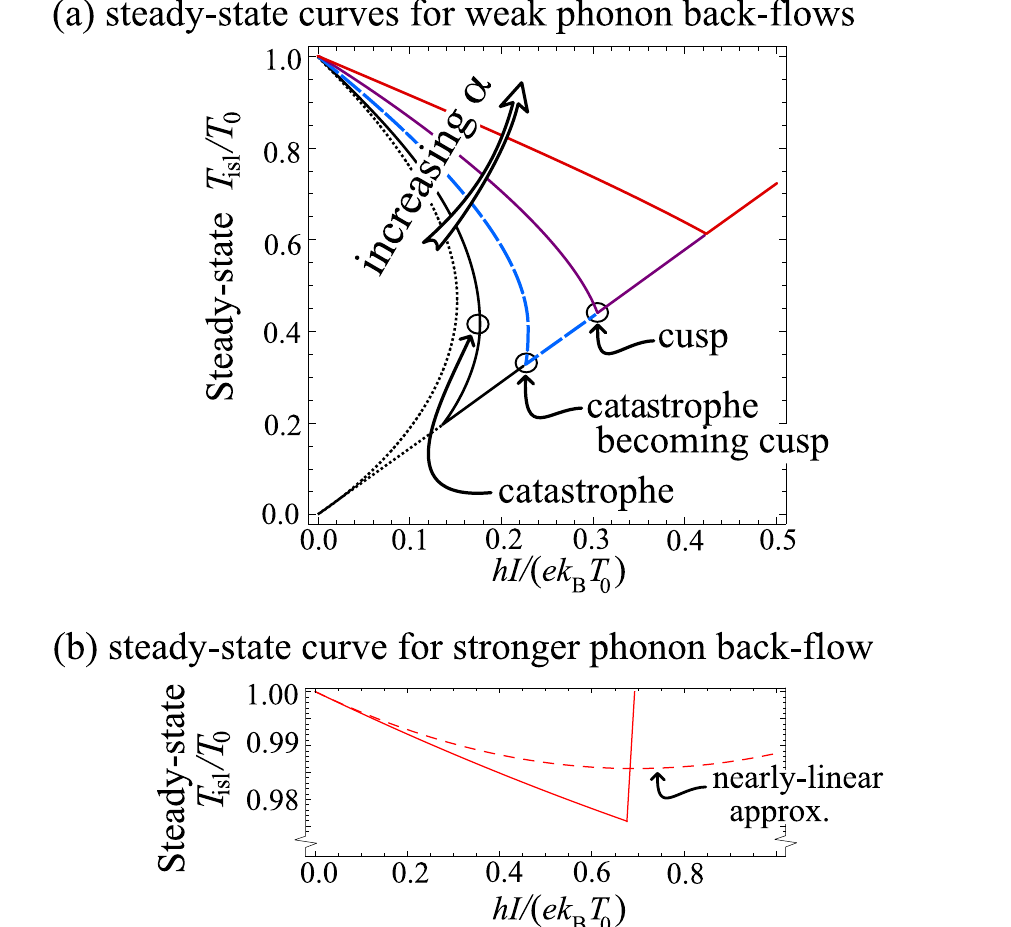}
\caption{\label{Fig:phonons}
(a) Steady-state refrigeration curves, $J(T_{\rm isl},I) = 0$ ,
for increasing heat flow due to phonons and photons;
$\alpha/\alpha_0  = 0, 0.02, 0.06, 0.12, 0.3$.
(b) The solid v-shaped curve is the steady-state refrigeration curve, 
$J(T_{\rm isl},I) = 0$ for $\alpha=10\alpha_0$, 
i.e.\ about 30 times stronger phonon back-flow than (a)'s upper-most curve. 
It is very different from the nearly-linear theory 
for the same $\alpha$ (dashed parabola).  
}
\end{figure}

\section{Refrigeration with phonons or photons}

I assume 
the metallic island is a suspended nanostructure\cite{suspended-structures}
in a cryostat at 0.3K, coupled to the substrate by suspended nanowires carrying the wires forming 
the cooling circuit.  
Wien's displacement law gives a photon wavelength of 10mm at 0.3K.
For any island smaller than a few millimetre, bulk blackbody radiation is replaced by a single photon 
mode of noise-flow between the island and its environment along the wires \cite{Pendry1983,Schmidt2004}, 
with $J_{\rm ph}=  r\alpha_0 (T_0^2-T_{\rm isl}^2)$,
where 
$\alpha_0 = \pi^2 \kB^2/(6h)$ is the ``quantum'' of heat-flow 
and $r$ is the mode's transmission.
If the environment part of the circuit has an inductance\cite{Hekking:photonic} $>1 \mu$H 
(or its capacitor equivalent), then  $r \ll 1$.

For a nanowire with $N_{\rm ph}$ phonon (vibrational) modes,
$J_{\rm ph}= \alpha (T_0^2-T_{\rm isl}^2)$, with $\alpha=N_{\rm ph} {\cal T}\alpha_0$, with  average transmission per mode of ${\cal T}$.
Experimental nanowires \cite{suspended-structures} show 
$\alpha \sim 0.3\alpha_0$ at ambient temperature $T_0=0.3K$. For this $\alpha$,
the steady-state curve has a pronounced cusp (uppermost v-shaped curve in 
Fig.~\ref{Fig:phonons}a), very different from the parabola given by the standard nearly-linear
theory. 
Fig.~\ref{Fig:phonons}b shows that this cusp persists up to such larger $\alpha$ 
that there is very little refrigeration (note the vertical axis shows $T_{\rm isl}/T_0$ is only slightly below one for any $I$).
The $\alpha$ chosen for the plot in Fig.~\ref{Fig:phonons}b corresponds to that 
observed in experiments in Ref.~[\onlinecite{suspended-structures},\onlinecite{suspended-structures2}] at 3K.
This is about thirty times larger than the minimum 
phonon conductance observed in Ref.~[\onlinecite{suspended-structures}]  at $0.3\,$K.
It also shows that the nearly-linear theory (dashed parabola) significantly 
under-estimates the optimum refrigeration.

Taking  the longitudinal phonon modes (velocity
9000ms$^{-1}$) and three types of transverse modes (velocity 6000ms$^{-1}$), 
the above experimental nanowires \cite{suspended-structures2} (with cross-section 200nm$\times$100nm)
have $N_{\rm ph} \sim 20$ at $T_0 \sim 0.3\,$K.
Evidently ${\cal T} \sim 1/60$, its smallness is probably due to the
frequency mis-match
between the phonons in the nanowire and the bulk.
If wires with cross-section 50nm$\times$50nm could be made,
then $N_{\rm ph} \sim 4$. Thus
$\alpha \sim 0.06\alpha_0$ can be expected (i.e.\ five times smaller than the current nanowires), 
the dashed curve in Fig.~\ref{Fig:phonons}a show the catastrophe emerging
at such $\alpha$.
To reduce $\alpha$ further, one can add surface roughness
or serpentines \cite{suspended-structures2}.

\begin{figure}
\includegraphics[width=\columnwidth]{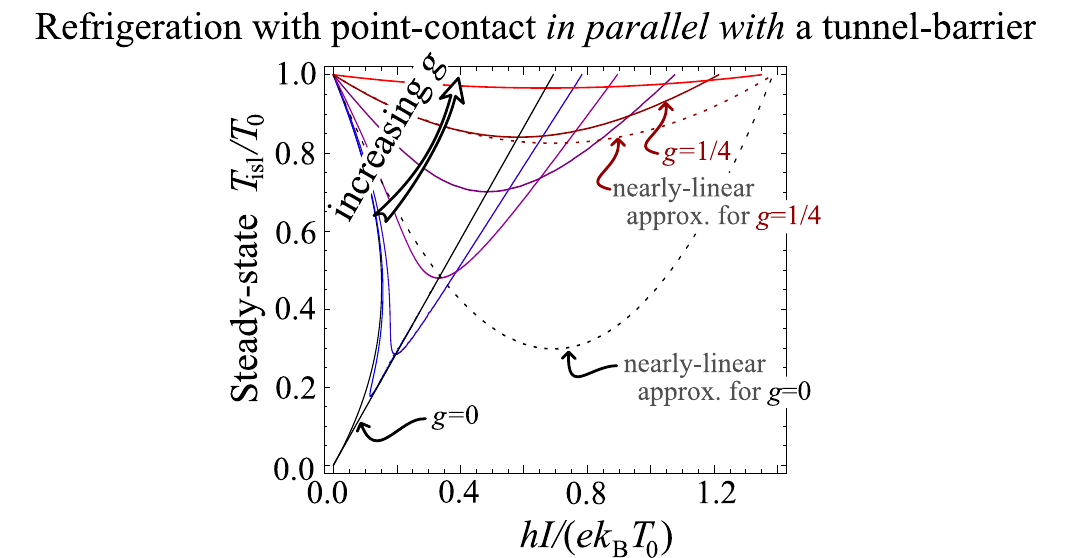}
\caption{\label{Fig:composite-system}
Steady-state refrigeration curves
for a composite system; point-contact in parallel with a barrier of conductance $G_{\rm barrier}= e^2g/h$ 
for $g=0,0.001,0.005,0.025,0.1,0.25,1$  (solid curves).
The parabolas (dashed) are the nearly-linear approximation
for $g=0,1/4$; for  $g=1$ the difference from the exact curve is not visible.}
\end{figure}

\section{Cross-over for point-contact in parallel with barrier}
\label{sect:composite}

I ask how one can induce a transition to the parabolic behavior in
Eq.~(\ref{Eq:J-nearly-linear}), since phonons, etc, do not do so?
I find that a transition only occurs upon  
reducing the $\eps$-dependence of ${\cal A}_{ij}$, 
lowering the ratio of the thermoelectric response to the usual electric response ---
for example, replacing the point-contact with a composite system consisting of 
a point-contact in {\it parallel} with a tunnel-barrier whose transmission is $\eps$-independent.
Fig.~\ref{Fig:composite-system} shows the steady-state response of 
the composite system,
for barrier conductance $G_{\rm barrier}=e^2g/h$. 
Upon increasing $g$ from zero, 
a transition occurs at 
$g=g_{\rm c} \sim 1/200$;
for $g > g_{\rm c}$,  the curve is single-valued, 
so $T_{\rm min}$ becomes a continuous function of $I$.
The nearly-linear theory works for $g\gtrsim 1$, 
deviations are still visible for $g=1/4$ 
(cf. solid and dashed curves). This fits the argument in Section~\ref{Sect:breakdown}, 
since the composite system has 
$\widetilde{\Pi} < \half G^{-1}$ for $g > (3\ln[2]-1)/2 \simeq 0.53$.

\begin{figure}
\includegraphics[width=\columnwidth]{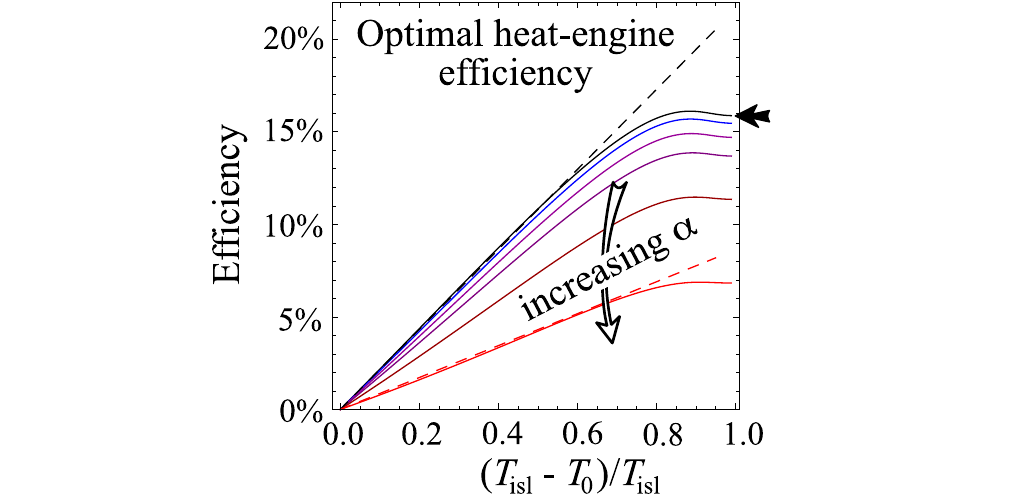}
\caption{\label{Fig:heatengine}
Heat-engine efficiency curves for 
$\alpha/\alpha_0 = 0, 0.02,0.06, 0.12,0.3, 1$.
Dashed lines are the linear-response predictions, Eq.~(\ref{Eq:Eff-linear}), for $\alpha/\alpha_0 = 0, 1$.
}
\end{figure}

\section{Heat-engine efficiency}
Returning to the case of a point-contact alone (without a tunnel-barrier),
I now consider its maximum efficiency as a heat engine. 
For  $T_{\rm isl}>T_0$, the circuit in Fig.~\ref{Fig:circuit}a 
provides electrical power $P=IV$ to any load connected between L and R. 
To calculate the maximum electrical power $P(T_{\rm isl},I)$ that a heat engine can extract from a heat flow 
$J(T_{\rm isl},I)$, one assumes a Ohmic load --- so $V(T_{\rm isl},I) = I/G_{\rm load}$ ---
is connected across its terminals, and adjust $G_{\rm load}$ to optimize the ratio of the power
at the load $P(T_{\rm isl},I)=I\,V(T_{\rm isl},I)$ to the heat flow $J(T_{\rm isl},I)$.
This corresponds to finding the $I=I_{\rm opt}$ which maximizes $P(T_{\rm isl},I)/J(T_{\rm isl},I)$.  
Maxima and minima are given by $P'J = PJ'$ where the primed is $(\rmd / \rmd I)$.
Given $V(T_{\rm isl},I)$ and $J(T_{\rm isl},I)$,  one can solve this to find $I_{\rm opt}$.
Optimal efficiency is  $\eta =P(I_{\rm opt})/J(I_{\rm opt})$.

As a warm-up, I consider the usual linear problem, with
$V(T_{\rm isl},I) =  S\,(T_{\rm isl}-T_0)\,-\,G^{-1} \, I $ and
$J(T_{\rm isl},I) =  \Theta \,(T_{\rm isl}-T_0) \,+\,  \Pi \,I $,
with $\Pi=\Gamma/G$ and $S=B/G$.
Calculate the optimal efficiency in the manner described above, I find
$I_{\rm opt}= ( \Theta/\Pi)[\sqrt{Z(T_{\rm isl})T_{\rm isl}+1} -1 ] \,(T_{\rm isl}-T_0)$.
Dropping $T$-dependences of $ZT$, one gets Eq.~(\ref{Eq:Eff-linear}).

Now I use the same method to get the efficiency in the nonlinear regime.
An analytic solution of $P'J=PJ'$ can be found for large  $(T_{\rm isl}/T_0)$,  
using the fact (confirmed by the numerics) that in this limit 
$-e V_1(I_{\rm opt})/(\kB T_0) \gg 1$  with $V_1<0$.  Otherwise the solution must be found numerically (see below).
For large  $(T_{\rm isl}/T_0)$, I take $\ln[1+\e^\mu] \to \mu$ 
and  ${\rm Li}_2 (-\e^{\mu}) \to  - \half \mu^2$ for large $\mu$, and find
$eV_1(T_{\rm isl},I)\big/(\kB T_0) \simeq   -t_{\rm isl} \ln[2] + \tilde I$
with
$
h J_1(T_{\rm isl},I)\big/ (\kB T_0)^2 \simeq  -\kappa t_{\rm isl}^2/2  - \ln[2] \,\tilde I \,t_{\rm isl} 
+ {\tilde I}^2/2$, 
%
where I define $t_{\rm isl}= T_{\rm isl}/T_0$, $\tilde I=hI \big/ (e \kB T_0)$ and $\kappa= \pi^2/6 - \ln^2[2] \simeq 1.16$.
The heat-current from the hot source into the device is $J(T_{\rm isl},I)=-J_1(T_{\rm isl},I)$,
and $P= -V_1(T_{\rm isl},I) I$ (given that $V_1 <0$).
In this case 
$P'(T_{\rm isl},I_{\rm opt})J(T_{\rm isl},I_{\rm opt}) = P(T_{\rm isl},I_{\rm opt})J'(T_{\rm isl},I_{\rm opt})$ 
is a quadratic equation for $I_{\rm opt}$; solving it gives
\begin{eqnarray}
{h I_{\rm opt} \over e\kB T_0} = {\kappa \over \ln[2] } 
\left[\sqrt{1+ \ln^2[2]/ \kappa} \ -1 \right] \, 
{T_{\rm isl} \over T_0},
\nonumber 
\end{eqnarray}
Without phonons or photons, the optimal efficiency tends to 
$1-\sqrt{1- 6 (\ln[2]/\pi)^2} \simeq 15.9\%$  for $T_{\rm isl}\to \infty$ 
(solid arrow in Fig.~\ref{Fig:heatengine}).
Solving $P'J=PJ'$ with Eqs.~(\ref{Eq:IvsV},\ref{Eq:J1-vs-I}) numerically,  
to find $I_{\rm opt}$ for different $T_{\rm isl}/T_0$, 
I plot $\eta$ against $(T_{\rm isl}-T_0)/T_{\rm isl}$ in Fig.~\ref{Fig:heatengine}.
I have no simple argument why the curves are 
slightly peaked at $(T_{\rm isl}-T_0)/T_{\rm isl} \sim 0.85$.

\begin{figure}
\includegraphics[width=\columnwidth]{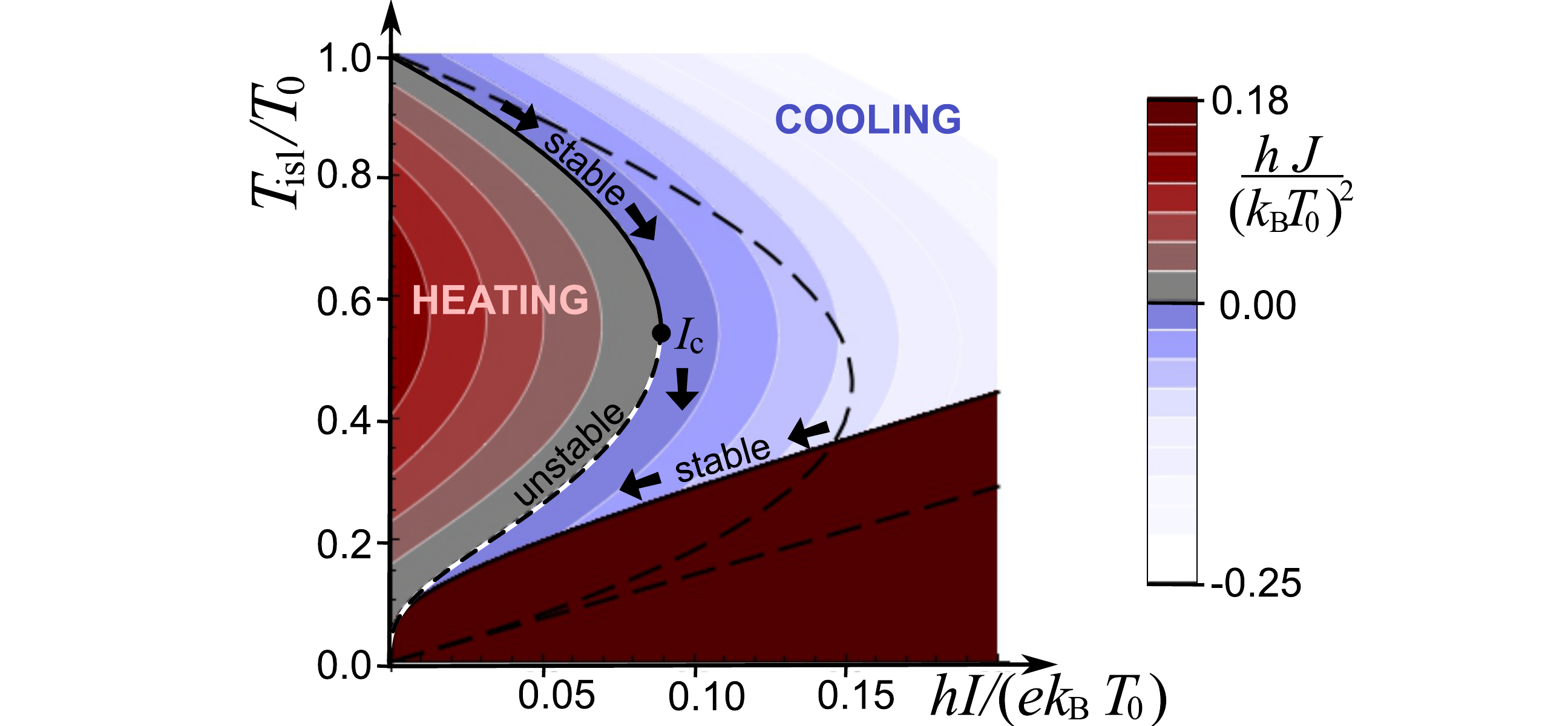}
\caption{\label{Fig:offset}
The heat-current when the barrier peak is at $E_{\rm pc}= \kB T_0/4$.
The results which give this curve will be discussed elsewhere.
For comparison, the thin-dashed curves are the steady-state for $E_{\rm pc}= 0$
in Fig.~1.
}
\end{figure}

\section{Catastrophe away from pinch-off}
Fig.~\ref{Fig:offset} gives an example showing the catastrophe is still present 
when the point-contact is away from pinch-off; i.e.\ when the barrier's peak is above the 
chemical potential of the island in Fig.~\ref{Fig:Cooling-no-phonons}. 
The catastrophe is present when $0< E_{\rm pc} \lesssim \kB T_0$, which corresponds 
to the parameter regime where a significant 
thermoelectric response was found experimentally in Ref. [\onlinecite{Molenkamp1992}].
The formulas leading to Fig.~\ref{Fig:offset} are similar to Eqs.~(\ref{Eq:IvsV}-\ref{Eq:J1-vs-I}) but rather longer, 
so I do not give them here.

\section{Concluding remarks}
I have shown that the point-contact (arguably the simplest thermoelectric nanostructure)
has a rich nonlinear behavior. 
In particular when it is used as a refrigerator, it exhibits multiple steady-states (stable and unstable)
and a fold catastrophe; or a sharp-cusp when there is significant phonon back-flow.  
I see no reason to think that more complicated nanoscale thermoelectric systems 
\cite{Casati2008,Nozaki2010,Saha2011,Wierzbicki2011,Karlstrom2011,Gunst2011,Rajput2011,Trocha2012}
have less rich behaviors.
Indeed a large $ZT$ is a strong hint that its
nonlinear Peltier term, $\widetilde{\Pi}I^2$
may dominate over its Joule heating term $-\half G^{-1} I^2$.
Section~\ref{Sect:breakdown} then gives a simple argument that $ZT$ ceases to give even a qualitative indication of how good a refrigerator it is. 
Thus the fully nonlinear response of such systems require detailed study,
beyond the weak nonlinearities considered in 
Refs.~[\onlinecite{Cipiloglu2004,Zebarjadi2007,Meair-Jacquod2013,Lopez-Sanchez2013}].

Finally, I recall that this work considered the case where the charging effects of electrons
at the point contact were well screened by the gates (or could be compensated for by the gates), meaning the $E_{\rm pc}$ in Fig.~\ref{Fig:circuit}
does not significantly change with bias.
Elsewhere, I will show that qualitatively similar effects can occur for point-contacts 
(and other systems) without gates, for which $E_{\rm pc}$ depends on the bias.

Since the submission of this work, a number of closely related works have appeared
\cite{Meair-Jacquod2013,Lopez-Sanchez2013,Jordan-et-al-2013,
Hwang-Sanchez-Lee-Lopez2013,w2013-best,Hershfield2013}.

\subsection{Acknowledgements} 
I thank F.\ Hekking, G.\ Rastelli, C.\ Stafford,
and particularly Ph.\ Jacquod, for insightful comments at various stages of this work.

\appendix
\section{Example of a self-consistent solution}

Here is a simple model of the point-contact for which 
the self-consistent solution can be found easily. However, the results 
in the body of the manuscript apply for almost  any self-consistent model. 
The point-contact is treated as a one-dimensional scattering problem (along the x-axis), 
see Fig.~\ref{Fig:circuit}d. 
Close  to the point-contact, this takes the form
$qV(x) = E_{\rm g} - \kappa x^2 +qV_{\rm scr}(x-x_{\rm pc})$
with energy measured from the island's chemical potential. 
The transverse confinement induces the $(E_{\rm g} -\kappa x^2)$-term,
where $E_{\rm g}$ can be tuned, since it equals $eV_{\rm g}$ minus a 
geometry-dependent constant. 
The $qV_{\rm scr}$-term is screening inside the electron gas, 
which I take as
\begin{eqnarray}
V_{\rm scr}(x) =  \left\{ \begin{array}{ccc} 
a V_{\rm L} & \hbox{for } & \ x \, < - l_{\rm scr}\\
a V_{\rm L} (l_{\rm scr}- x) \big/(2l_{\rm scr})
 & \hbox{for }  & |x| \leq \  l_{\rm scr}\ \\
0 &  \hbox{for } & x \, > \ l_{\rm scr}
\end{array} \right.
\nonumber 
\end{eqnarray}
with $x_{\rm pc}$ being the self-consistently determined peak of $qV(x)$.
A little algebra gives $x_{\rm pc}= -a qV_{\rm L}/(4\kappa l_{\rm scr})$,
thus the energy at the peak is $E_{\rm pc}=qV(x_{\rm pc}) 
=E_{\rm g}+\half a qV_{\rm L}\left(1- a q V_{\rm L}/(8\kappa l_{\rm scr}^2) \right)$.
Finally I note that both $a$ and $l_{\rm scr}$ depend on the scattering matrix of the junction, which
in turn depends on $E_{\rm pc}$.  To solve this problem self-consistently, I assume one is in the
regime where $e_{\rm pc}=E_{\rm pc}-E_{\rm g}$ is small enough to 
approximate 
$a= a_0(1+b_a \, e_{\rm pc})$
and 
$l_{\rm scr}=  l_{\rm scr0}(1+b_l e_{\rm pc})$. 
If necessary, $a_0,l_{\rm scr0},b_a,b_l$ can be found by simulating Poisson's equation;
typically $e_{\rm pc}$ is small for small $a$.
Then $E_{\rm pc}$ is equal to a linear function of itself;
re-arranging this gives
\begin{eqnarray}
E_{\rm pc} -  E_{\rm g} = { a_0 qV_{\rm L}/2 - C(qV_{\rm L}) 
\over 1- a_0 b_a qV_{\rm L} + 2C(qV_{\rm L}) \, 
\big[ b_a - b_l \big] },
\nonumber
\end{eqnarray}
where I define
$C(qV_{\rm L})  = \big(a_0 qV_{\rm L}/l_{\rm scr0} \big)^2 \big/(16 \kappa)$.
Thus the right hand side of this equation is the $E_{\rm scr}(aqV_{\rm L})$ mentioned in the body of the text.
As mentioned earlier, I assume that
tunnelling at energies $\eps < E_{\rm pc}$ is negligible,  
so ${\cal A}_{\rm LM}(\eps-E_{\rm pc}>0)=-1$ and  ${\cal A}_{\rm LM}(\eps-E_{\rm pc}<0)=0$.
To see that this respects gauge-invariance, I recall that $\eps, E_{\rm pc},V_{\rm L,g}$
are all measured relative to the island's potential, and replace them by quantities  
measured from a fixed external reference,
so the island is at $\widetilde{V}_{\rm M}$.
For clarity, here (unlike in the paragraph containing Eq.~(\ref{Eq:J-initial})) 
it is necessary to use a tilde to explicitly indicate quantities measured from the external reference.
I make the replacement
$qV_{\rm L}= (\widetilde{\eps}-q\widetilde{V}_{\rm M})-(\widetilde{\eps}-q\widetilde{V}_{\rm L})$. 
From this, one sees that ${\cal A}_{\rm LM}(\eps-E_{\rm pc})$
is only a function of the set of differences 
$\{\widetilde{\eps}-q\widetilde{V}_k\}$, and so respects gauge-invariance.



\newpage
\appendix



\begin{thebibliography}{1}



\bibitem{books}
H.J.~Goldsmid, {\it Thermoelectric Refrigeration} (Temple Press, London, 1964),
chapt 1.
H.J.~Goldsmid, {\it Introduction to Thermoelectricity} (Springer, Heidelberg, 2009),
chapt 2. 
\bibitem{DiSalvo-review}
F.J.~DiSalvo, Science {\bf 285},703 (1999).
\bibitem{Shakouri-reviews}
A.~Shakouri and M.~Zebarjadi 
Chapt 9 of {\it Thermal nanosystems and nanomaterials}, S.~Volz (Ed.)  (Springer, Heidelberg, 2009).
A.~Shakouri , Annu. Rev. Mater. Res. {\bf 41}, 399 (2011).



\bibitem{Casati2008}
G.~Casati, C.~Mej\'ia-Monasterio, and T.~Prosen,
Phys.~Rev.~Lett.~{\bf 101}, 016601 (2008).
\bibitem{Nozaki2010}
D.~Nozaki, H.~Sevin\c cli, W.~Li, R.~Guti\'errez, and G.~Cuniberti,  Phys.~Rev.~B {\bf 81}, 235406 (2010). 
\bibitem{Saha2011}
K.K.~Saha, T.~Markussen, K.S.~Thygesen, and B.K.~Nikoli\'c, Phys.~Rev.~B {\bf 84}, 041412(R) (2011).
\bibitem{Wierzbicki2011}
M.~Wierzbicki and R.~Swirkowicz, Phys.~Rev.~B {\bf 84}, 075410 (2011). 
\bibitem{Karlstrom2011}
O. Karlstr\"om, H. Linke, G. Karlstr\"om, and A. Wacker,  Phys.~Rev..~B {\bf 84}, 113415 (2011). 
\bibitem{Gunst2011}
T.~Gunst, T.~Markussen, A.-P.~Jauho, and M.~Brandbyge, Phys.~Rev.~B {\bf 84}, 155449 (2011).
\bibitem{Rajput2011}
G.~Rajput, and K.C.~Sharma,  J.~Appl.~Phys.~{\bf 110}, 113723 (2011).
\bibitem{Trocha2012}
P.~Trocha and J.~Barna\'s,  Phys.~Rev.~B {\bf 85}, 085408 (2012).





\bibitem{Pekola-reviews}
F.~Giazotto, T.T.~Heikkila, A.~Luukanen, A.M.~Savin, J.P.~Pekola,
Rev.~Mod.~Phys.~{\bf 78}, 217 (2006).
J.T.~Muhonen, M.~Meschke, J.P.~Pekola,  Rep.~Prog.~Phys.~{\bf 75}, 046501 (2012).


\bibitem{SC-cooling-expt1} 
M.M.~Leivo, J.P.~Pekola and D.V.~Averin, Appl.\ Phys.\ Lett.\ {\bf 68}, 1996 (1996).


\bibitem{SC-cooling-expt2} 
A.M.\ Clark, N.A.\ Miller, A.\ Williams, S.T.\ Ruggiero, G.C.\ Hilton, L.R.\ Vale, J.A.\ Beall, 
K.D.\ Irwin, and J. N.\ Ullom,
Appl.\ Phys.\ Lett.\ {\bf 86}, 173508 (2005).

\bibitem{SC-cooling-expt3} 
S.~Rajauria, P.S.~Luo, T.~Fournier, F.W.J.~Hekking, H.~Courtois, and B.~Pannetier, 
Phys.~Rev.~Lett.~{\bf 99}, 047004 (2007).
S.~Rajauria , P.~Gandit , F.W.J.~Hekking, B.~Pannetier , H.~Courtois, 
J.~Low Temp.~Phys.~{\bf 154} 211 (2009).


\bibitem{Footnote:semicond}
For bulk semiconductors, linear response (Boltzmann) transport theory
works when the temperature difference on the scale of the inelastic scattering length
is small compared with the average temperature at that point.  
The inelastic scattering length (typically 1-100nm at 290K) is very much less
than the length over which temperature drops (length of semiconductor, e.g. millimetres).
Then a linear theory works well even when the temperature drop is large, although
nonlinear corrections have been studied \cite{Zebarjadi2007}.
In contrast, nanostructures are often smaller than the inelastic scattering length
(often much more than 1$\mu$m at 1K).
Ideally the whole temperature drop occurs across this nanostructure, making 
linear-response theory inappropriate when the temperature drop is significant.

\bibitem{Zebarjadi2007}
M.~Zebarjadi, K.~Esfarjani, and A.~Shakouri, Appl. Phys. Lett. {\bf 91}, 122104 (2007);
in {\it Thermoelectric Power Generation} by T.P. Hogan, J. Yang, R. Funahashi, T. Tritt (Eds.), MRS Symposia Proceedings No.\ {\bf 1044}, (Materials Research Society, Pittsburgh,2008) p.\, U10. 



\bibitem{Molenkamp1992} 
L.W.~Molenkamp, Th.~Gravier, H.~ van Houten, O.J.A.~Buijk, M.A.A.~Mabesoone, and C.T.~Foxon,
Phys.~Rev.~Lett.~{\bf 68}, 3765 (1992).
H.~ van Houten, L.W.~Molenkamp, C.W.J.~Beenakker, and C.T.~Foxon,
Semicond.~Sci.~Technol.~{\bf 7}, B215 (1992).
\bibitem{Ghoshal2002}
U.~Ghoshal, S.~Ghoshal, C.~McDowell, L.~Shi, S.~Cordes, and M.~Farinelli,
Appl.~Phys.~Lett.~{\bf 80}, 3006 (2002)


\bibitem{Bogachek1998}
E.N.~Bogachek, A.G.~Scherbakov and U.~Landman,
Solid State Comm.~{\bf 108}, 851 (1998). 
This work calculates the non-linear differential Peltier coefficient 
$\Pi_{\rm diff} = \rmd J /\rmd I$, even though some formulas assume linear behavior.
\bibitem{Cipiloglu2004}
M.A.~\c Cipilo\v glu, S.~Turgut and M.~Tomak,
Phys.~Stat.~Sol.~(b) {\bf 241}, 2575 (2004).
\bibitem{Nakpathomkun-Hu-Linke2010}
N.~Nakpathomkun, H.Q.~Xu, and H.~Linke,
Phys.~Rev.~B {\bf 82} 235428 (2010).








\bibitem{Moskalets1995}
M.~Moskalets, JETP Lett. 62, 719 (1995).

\bibitem{Christen-ButtikerEPL96}
T.~Christen and M.~B\"uttiker, Europhys.~Lett.~{\bf 35}, 523 (1996). 

\bibitem{Sanchez-Buttiker}
D.~Sanchez and M.~Buttiker, Phys.~Rev.~Lett.~{\bf 93}, 106802 (2004).





\bibitem{Meair-Jacquod2012}
J.~Meair and Ph.~Jacquod,  J.~Phys.: Condens.\ Matter {\bf 24}, 272201  (2012).


\bibitem{Sanchez-Lopez2012}
D.~S\'anchez, and R.~L\'opez,  Phys.\ Rev.\ Lett.\ {\bf 110}, 026804 (2013).






\bibitem{Butcher1990}
P.N.~Butcher, J.~Phys.: Condens.~Matter,  {\bf 2}, 4869 (1990). 



\bibitem{suspended-structures}
J.S.~Heron, T.~Fournier, N.~Mingo, O.~Bourgeois,	
Nano Lett.\ {\bf 9},  1861 (2009).


\bibitem{suspended-structures2}
J.-S.\ Heron, C.\ Bera, T.\ Fournier, N.\ Mingo, and O.\ Bourgeois,
Phys.~Rev.~B 82, 155458 (2010).

\bibitem{Gasparinetti2011}
S.\ Gasparinetti, F.\ Deon, G.\ Biasiol, L.\ Sorba, F.\ Beltram, and F.\ Giazotto,
Phys.\ Rev.\ B {\bf 83}, 201306(R) (2011).


\bibitem{Engquist-Anderson1981}
H.-L.~Engquist and P.W.~Anderson, Phys.~Rev.~B {\bf 24}, 1151 (1981).
\bibitem{Sivan-Imry1986}
U.~Sivan and Y.~Imry,  Phys.~Rev.~B {\bf 33}, 551 (1986).
\bibitem{Claughton-Lambert}
N.R.~Claughton and C.J.~Lambert, Phys.~Rev.~B {\bf 53}, 6605 (1996).
\bibitem{jw-epl}
Ph.~Jacquod and R.S.~Whitney, Europhys.~Lett.~{\bf 91}, 67009 (2010).


\bibitem{Liu2011}
Y.-S.~Liu, B.C.~Hsu, and Y.C.~Chen,
J.~Phys.~Chem.~C {\bf 115}, 6111 (2011). 

\bibitem{jwmb-onsager}
Ph.~Jacquod, R.S.~Whitney, J.~Meair, and M.~B\"uttiker,  Phys.\ Rev.\ B {\bf 86}, 155118 (2012).


\bibitem{footnote:Theta}
$\Theta_i$ is the thermal conductance for $I=0$
(not $V=0$).


\bibitem{footnote:ferro-para}
In $\phi^4$-theory at high temperatures, the $\phi^2$-term is positive and determines the minimum of free-energy. At low temperatures,
the $\phi^2$ term is negative and the minimum is determined by a higher 
order term (the  $\phi^4$-term).
In the refrigerator, the role of $\phi$ is played by $I$, while the parameter
which controls the sign of the quadratic term is $\widetilde{\Pi}$ rather than temperature.


\bibitem{Entin-Wohlman2010}
O.~Entin-Wohlman, Y.~Imry, and A.~Aharony,
Phys.\ Rev.~B {\bf 82}, 115314 (2010).

\bibitem{SSJB2012}
B.~Sothmann, R.~S\'anchez, A.N.~Jordan, and M.~B\"uttiker,
Phys.~Rev.~B 85, 205301 (2012).

\bibitem{Sothmann-Buttiker2012}
B.~Sothmann, and M.~B\"uttiker, EPL 99, 27001 (2012).

\bibitem{Sanchez-Buttiker2012}
R.~S\'anchez and M.~B\"uttiker, EPL {\bf 100}, 47008 (2012).

\bibitem{Entin-Wohlman2013}
J.-H.\ Jiang, O.\ Entin-Wohlman, Y.\ Imry,
New J.\ Phys.\ {\bf 15}, 075021 (2013).
O.\ Entin-Wohlman, A.\ Aharony, Y.\ Imry, arXiv:1306.1813.






\bibitem{w-joule}
R.S.~Whitney, Phys.\ Rev.\ B {\bf 87}, 115404 (2013).

\bibitem{Buttiker1993-96}
M.~B\"uttiker, J.~Phys.\ Condens.\ Matter, {\bf 5}, 9361 (1993).
M.~B\"uttiker, A.~Pr\^etre, and H.~Thomas, Phys.~Rev.~Lett.\ {\bf 70}, 4114 (1993);
Z.~Phys.~B, {\bf 94}, 133 (1994).  
T.~Christen and M.~B\"uttiker, Phys.~Rev.~Lett.\ {\bf 77}, 143 (1996). 

\bibitem{Petitjean2009}
C.~Petitjean, D.~Waltner, J.~Kuipers, I.~Adagideli, K.~Richter,
Phys.~Rev.~B, {\bf 80}, 115310 (2009).









\bibitem{footnote:tunnelling}
The results are unchanged if a little bit of tunnelling means that
${\cal A}_{\rm LM}(\eps)$  goes smoothly from $0$ to $-1$ on a scale less than $\kB T_{\rm isl}$.







\bibitem{Pendry1983}
J.B.~Pendry, J.\ Phys.\ A:Math.\ Gen.\  {\bf 16}, 2161 (1983).

\bibitem{Schmidt2004}
D.R.\ Schmidt, R.J.\ Schoelkopf, and A.N.\ Cleland, 
Phys.\ Rev.\ Lett.\ {\bf 93}, 045901 (2004). 

\bibitem{Hekking:photonic}
L.M.A.~Pascal, H.~ Courtois, and F.W.J.~Hekking,
Phys.\ Rev.~B {\bf 83}, 125113 (2011).






 

\bibitem{Meair-Jacquod2013}
J.~Meair, and Ph.~Jacquod,  J.~Phys.: Condens.\ Matter, {\bf 25}, 082201  (2013). 
\bibitem{Lopez-Sanchez2013}
R.~L\'opez, and D.~S\'anchez,
Phys.\ Rev.\ B {\bf 88}, 045129 (2013).

\bibitem{Jordan-et-al-2013}
A.N.\ Jordan, B.\ Sothmann, R.\ S\'anchez, and M.\ B\"uttiker
Phys.\ Rev.\ B {\bf 87}, 075312 (2013).

\bibitem{Hwang-Sanchez-Lee-Lopez2013}
S.-Y.\ Hwang, D.\ S\'anchez, M.\ Lee, and R.\ L\'opez,
arXiv:1306.6558.

\bibitem{w2013-best}
R.S.\ Whitney, arXiv:1306.0826.
\bibitem{Hershfield2013}
S.\ Hershfield, K.A.\ Muttalib, B.J.\ Nartowt,
arXiv:1307.5670.

\end{thebibliography}
\end{document}